\documentclass[11pt]{article}
\usepackage{fancyhdr,graphicx}
\usepackage{latexsym,amssymb,epsfig}
\usepackage{amsmath,amssymb}
\usepackage{caption2}
\usepackage{psfrag}

\newcommand{\be}{\begin{equation}}
\newcommand{\ee}{\end{equation}}
\newcommand{\bea}{\begin{eqnarray}}
\newcommand{\eea}{\end{eqnarray}}

\newcommand{\fr}{\frac}

\newcommand{\Pcal}{{\cal{P}}}
\newcommand{\Qcal}{{\cal{Q}}}

\input epsf

\begin{document}

\begin{titlepage}

\begin{flushright}
\end{flushright}
\vspace*{1.5cm}
\begin{center}
{\Large \bf Towards understanding Regge trajectories in holographic\\\vskip 0.2cm QCD}\\[2.0cm]

{\bf Oscar Cat\`{a}\footnote{Fulbright fellow. Formerly at Grup de F{\'\i}sica Te{\`o}rica and IFAE, Universitat
Aut{\`o}noma de Barcelona, 08193 Barcelona, Spain.
}}\\

{\it{Department of Physics, University of
Washington,  Seattle, WA 98195}}\\

{\it{E-mail:}} {\textrm{ocata@phys.washington.edu}}

\end{center}

\vspace*{1.0cm}

\begin{abstract}

We reassess a work done by Migdal on the spectrum of low-energy vector mesons in QCD in the light of the AdS-QCD correspondence. Recently, a tantalizing parallelism was suggested between Migdal's work and a family of holographic duals of QCD. Despite the intriguing similarities, both approaches face a major drawback: the spectrum is in conflict with well-tested Regge scaling. However, it has recently been shown that holographic duals can be modified to accomodate Regge behavior. Therefore, it is interesting to understand whether Regge behavior can also be achieved in Migdal's approach. In this paper we investigate this issue. We find that Migdal's approach, which is based on a modified Pad\'e approximant, is closely related to the issue of quark-hadron duality breakdown in QCD.
\end{abstract}

\end{titlepage}

\section{Introduction}
The recent years have seen an emergence of 5-dimensional modelling in an effort to make contact with QCD, the main motivation under most of the models relying upon the gauge/gravity duality conjectured by Maldacena almost a decade ago.\footnote{See, for instance, \cite{Klebanov,Hoker} for comprehensive reviews.}

Even though the original Maldacena conjecture related a slice of superstring theory on anti-de Sitter (AdS) background with ${\cal{N}}=4$ SYM, both the belief that QCD is rooted in string theory and, especially, the high-energy conformal symmetry of QCD yield room for hoping that the conjecture may be extended beyond its actual formulation to provide new insights into the strong interactions. Indeed, rather simple models capturing some of the features of QCD like chiral symmetry breaking and confinement are in remarkable agreement with phenomenology. 

The common setting of these models is a 5-dimensional AdS background. A 4-dimensional boundary brane is introduced, upon which Standard Model fields are bound to propagate. Confinement is modelled by cutting the bulk space down at another boundary brane, thus introducing an infrared scale that mimics $\Lambda_{QCD}$. Spontaneous chiral symmetry breaking \cite{Katz-Son,Leandro,Hirn1}, quark masses \cite{Katz-Son,Leandro,Shock} and OPE condensates \cite{Hirn} have also been implemented over the last years.\footnote{See also, e.g., \cite{Ghoroku,Evans} for the effects of modifying the AdS metric in the infrared.} However, the manifestly wrong scaling of the predicted spectrum, namely $m_n^2\sim n^2$, has been a burden to all these models, not only from the phenomenological side but also because their eventual connection to string theory was doubtful. 

Only recently was it realized \cite{Karch} that the wrong mass-scaling was not an essential drawback of holographic models of QCD, but due to the choice of boundary conditions. One can draw an analogy between the 5-dimensional Kaluza-Klein modes in the bulk of the aforementioned models and the quantum mechanical excitations in a square well potential. By changing the profile of the boundaries one can dial the energy spectrum at will and should be able to get to the desired mass-scaling. In the language of holographic QCD, this amounts to a dynamical dilaton, scaling (asymptotically) as $\Phi(z)\sim z^2$, whereas the metric is kept (asymptotically) purely AdS. Therefore, hard-wall models, {\it{i.e.}}, those with a sharp infrared brane, yield $m_n^2\sim n^2$, as opposed to infrared-improved models, where the asymptotic form of the dilaton ensures $m_n^2\sim n$.

It turns out, as first pointed out in \cite{Shifman1}, that hard-wall holographic models bear a tantalizing resemblance with a work done by Migdal \cite{Migdal} almost 30 years ago on modelling the spectrum of vector mesons in QCD.\footnote{Recently it was also claimed \cite{Radyushkin} that the extension of Migdal's programme to three point functions carried out in \cite{Dosch} has its counterpart in the holographic light-front wavefunctions of \cite{Brodsky}.} This very ambitious project relied on linking short distance QCD to the meson spectrum through quark-hadron duality. The information on short distances was used to make a Pad\'e approximant to the whole correlator and thereby information on the spectrum was extracted. This procedure was designed to be a systematic step by step approximation, in the sense that adding more information on short distances one should get closer to the true QCD spectrum. Thus, at least formally, the programme was improvable. Therefore, when the Pad\'e for the leading order perturbation theory yielded a spectrum with poles at zeros of Bessel functions, meaning that the tower of vector mesons in QCD should scale as $m_n^2\sim n^2$, the apparent contradiction with experimental evidence was seen as an artifact of the truncation. Should more orders in the OPE be added, the spectrum would smoothly reshuffle to display Regge trajectories.    

The similarity between Migdal's result and that of hard-wall holographic models is, in a sense, not so surprising. In the simplest hard-wall models the 5-dimensional background is filled with a pure AdS metric, meaning that the correlator looks like that of a free quark theory. Both approaches therefore start from the same conformally invariant correlator. The interesting thing is that the analogy seems to go further and one might even argue \cite{Erlich} that holographic models can be interpreted as the limit of a series of Pad\'e approximants. A qualitative understanding of this should be found in the deconstructed version of holographic QCD models \cite{Son}, where an open moose model of hidden local symmetries \cite{Bando} was shown to be interpretable as the latticized version of a fifth dimension. A $[N,N]$ Pad\'e approximant would then be the dual picture to an open moose with $N$ hidden symmetries.

In \cite{Erlich}, Migdal's procedure was intended to shed some light on AdS-QCD. Here we reverse the point: we reassess Migdal's work with the benefit of using the geometrical intuition of AdS-QCD models. A natural question then arises: if Regge scaling was achieved in holographic models through a non-trivial dilaton background, what is the analog in Migdal's approach? We will see that one is forced to include non-perturbative effects beyond the OPE, generically referred to as {\it{quark-hadron duality violations}} in the QCD literature \cite{Shifman}. Actually, we will show that non-perturbative effects were already present in Migdal's original work. 

The discussion on quark-hadron duality breakdown will also allow us to grasp the meaning of the dilaton's quadratic scaling in holography, an interesting question partially answered in the context of string theory \cite{Karch}. Still, one would like to have some physical intuition in the context of QCD. 
We explicitly show that the dilaton's infrared profile in Regge-like holographic models is dual to a specific modeling of duality breakdown in QCD.  

The outline of the paper is as follows: sections 2, 3 and 4 deal with 4-dimensional models: in section 2 we introduce a toy model with explicit Regge scaling, while in section 3 we discuss Pad\'e approximants. Section 4 is devoted to an analysis of Migdal's modification of the Pad\'e approximant and the relationship with the issue of quark-hadron duality violations. We will work in the large-$N_c$ limit of QCD. There, resonances are narrow and therefore mass-poles are well-defined. The aim is to use our toy model, where results can be worked out analytically, as a probe to reassess the accuracy of Migdal's programme. Contact with 5-dimensional holographic models is made in the last section.

\section{A toy model}
Following \cite{Migdal}, we consider the following two-point vector correlator
\begin{equation}\label{def}
\Pi_V^{\mu\nu}(q)=i\int d^4x e^{iq\cdot x} \langle 0|T\{V^{\mu}_{{\bar{u}}d}(x)(V^{\nu}_{{\bar{u}}d})^{\dagger}(0)\}|0\rangle\, ,
\end{equation}
with currents defined as
\begin{equation}\label{current}
V^{\mu}_{{\bar{u}}d}(x)={\bar{u}}(x) \gamma^{\mu} d(x)\,.
\end{equation}
In the chiral limit, Lorentz symmetry implies that 
\begin{equation}\label{lorentz}
\Pi_V^{\mu\nu}(q)=(q^{\mu}q^{\nu}-q^2 g^{\mu\nu})\Pi_V(q^2)\, ,
\end{equation}
namely that there is only one form factor, which satisfies the following dispersion relation
\begin{equation}\label{altdisper}
\Pi_V(q^2)=q^2\int_0^{\infty}\fr{dt}{t(t-q^2-i\varepsilon)}\fr{1}{\pi}{\mathrm{Im}}\,\Pi_V(t)+\Pi_V(0)\, ,
\end{equation}
where $\Pi_V(0)$ is a subtraction constant.
In the strict large-$N_c$ limit, a two-point Green's function consists of single particle exchange of an infinite number of stable mesons \cite{Witten}. Therefore, the spectral function reads 
\begin{equation}\label{ansatz}
\fr{1}{\pi}{\mathrm{Im}}\,\,\Pi_V(t)=\sum_{n=0}^{\infty}F_n^2\delta(t-M_V^2(n))\, .
\end{equation}
Unfortunately, this is the furthest one can go in full generality. Decay constants and masses in large-$N_c$ QCD are unknown. However, phenomenology in $N_c=3$ QCD seems to indicate that Regge behavior holds and, in the vectorial channel, already sets in from the very first resonance. Therefore, we will choose as an educated ansatz \cite{Shifman}  
\begin{equation}\label{regge}
F_n^2=F^2,\qquad M_V^2(n)=m_V^2+an\, .
\end{equation}
Inserting ({\ref{regge}}) back into (\ref{altdisper}) we end up with the following expression for the correlator
\begin{equation}\label{green}
\Pi_V(q^2)=\fr{F^2}{a}\left[\psi\left(\fr{m_V^2}{a}\right)-\psi\left(\fr{-q^2+m_V^2}{a}\right)\right]\, ,
\end{equation}
where $\psi(\xi)$ is the digamma function, defined as\footnote{Notice the connection between this model and the Veneziano model (Beta function) for fixed spin, {\it{i.e.}},
\begin{equation}
B(n,s)=\fr{\Gamma{(n)}\,\,\Gamma{(s)}}{\Gamma{(n+s)}}\, .
\end{equation}
In both cases, the mass-poles are dictated by the poles of the Gamma function.}
\begin{equation}\label{digamma}
\psi(\xi)=\fr{d}{d\xi}\log{\Gamma(\xi)}\, .
\end{equation}
The free parameters $F$ and $a$ can now be fixed upon matching with the first terms in the OPE for $\Pi_V(Q^2\equiv -q^2)$. Indeed, in the Euclidean half-plane, the digamma function admits the following integral representation
\begin{equation}\label{Gauss}
\psi(\xi)=\int_0^{\infty} dt \left(\fr{e^{-t}}{t}-\fr{e^{-\xi t}}{1-e^{-t}}\right)\, ,
\end{equation}
from which the large-$\xi$ behavior can be extracted
\begin{equation}\label{dig}
\psi(\xi)=\log{\xi}-\fr{1}{2\xi}-\sum_{n=1}^{\infty}\fr{B_{2n}}{2n\,\xi^{2n}}\, .
\end{equation}
Thus, the OPE for our model reads
\begin{equation}\label{ope} \Pi_V^{OPE}(Q^2)=-\fr{F^2}{a}\log{\fr{Q^2}{\mu^2}}+\sum_{k=1}^{\infty}\fr{c_{2k}}{Q^{2k}}\, ,
\end{equation}
where the condensates are given by \cite{Cata05paper}
\begin{equation}\label{cond}
c_{2k}=(-1)^k\fr{F^2a^{k-1}}{k}\,B_k\left(\fr{m_V^2}{a}\right)\, ,
\end{equation}
and $B_k(\xi)$ stand for the Bernoulli polynomials.
On the other hand, in real QCD, to leading order in $\alpha_s$,
\begin{equation}
\Pi_V^{OPE}(Q^2)=-\fr{4}{3}\fr{N_c}{(4\pi)^2}\log{\fr{Q^2}{\mu^2}}+...
\end{equation}
Thus, we immediately get
\begin{equation}
\fr{4}{3}\fr{N_c}{(4\pi)^2}=\fr{F^2}{a}\, .
\end{equation}
As it stands, the model only fullfils one short-distance constraint. In principle, one could refine the model in order to match the first OPE condensates. This can be easily achieved by relaxing our initial ansatz for the decay constants (\ref{regge}) and letting an arbitrary number of them as free parameters. For such extensions of our model, we refer the reader to \cite{Cata05paper}. However, in this paper we will not be interested in the phenomenological implications of the toy model. For all our considerations we can safely stick to the simplest version of Eq.(\ref{green}). 

\section{Building the Pad\'e approximant}
With our model for the two-point vector current correlator of (\ref{green}), we can now attempt to build its Pad\'e approximant.\footnote{Pad\'e approximants have been a useful tool for widespread applications in condensed matter and particle physics. See, for instance, the very interesting examples of \cite{Gammel}.} With the advantage of knowing the final answer, we will try to assess the validity of the procedure followed in \cite{Migdal}. We begin with some notation and definitions. 

A $[N,M]$ Pad\'e approximant to a function is defined as the quotient of two polynomials ${\Pcal_M}$ and ${\Qcal_N}$
\begin{equation}\label{pade}
\Pi_V(q^2)\simeq\fr{\Pcal_M(q^2)}{\Qcal_N(q^2)}\, ,
\end{equation}
such that the first $N+M+1$ derivatives of the function at a point match those of the approximant. With full generality, we can set $\Qcal_N(0)=1$. Notice that one should be able to define (at least formally) a Taylor expansion of the function around a point, even if the expansion is divergent.
The next issue to address is that of convergence. One would wish that the $[N,M]$ rational should converge to the function as $N,M\rightarrow \infty$. This is true for a class of functions known as Stieltjes functions. Stieltjes functions admit the following power expansion about the origin
\begin{equation}\label{stieltjes}
f(z)=\sum_{n=0}^{\infty} f_n\,\, z^n\, ,
\end{equation}
with the coefficients (also called moments) given by the following integrals
\begin{equation}\label{moments}
f_n=(-1)^n\int_0^{\fr{1}{R}}u^n \nu(u)\, du\, ,
\end{equation}
where $R$ is the radius of convergence of (\ref{stieltjes}) and $\nu(u)$ is some positive definite integration measure. Therefore, (\ref{stieltjes}) admits the following representation as a Stieltjes integral
\begin{equation}
f(z)=\int_0^{\fr{1}{R}}\fr{du}{1+uz}\nu(u)\, .
\end{equation}
With the change of variable $u=t^{-1}$ one obtains
\begin{equation}\label{stielt}
f(z)=\int_R^{\infty}\fr{dt}{t(t+z)}\nu\left(\fr{1}{t}\right)\, ,
\end{equation}
which matches (\ref{altdisper}) with the following identifications\footnote{Notice that our ansatz for the spectrum (\ref{ansatz}) assumes a mass gap with the first resonance sitting at $q^2=m_V^2$. Therefore, $R=m_V^2$ in (\ref{stielt}). The spectral integral beginning at $t=0$ or $t=m_V^2$ thus yields the same result. We will hereafter use the former notation for simplicity.}
\begin{equation}
z=-q^2\, ,\qquad \nu\left(\fr{1}{t}\right)=\fr{1}{\pi}{\mathrm{Im}}\,\Pi_V(t)\, .
\end{equation}
Therefore, as recently pointed out in \cite{Santi}, the vector vacuum polarization admits a representation as a Stieltjes integral. This is most fortunate, and a series of interesting consequences ensue.\footnote{See, for instance, \cite{Baker}.} First and foremost, it ensures convergence of the Pad\'e approximant of (\ref{pade}) for the special subset $[(M+J),M]$, $J\geq -1$. The convergence region covers the whole complex $q^2$ plane, save for the Minkowski axis, where the poles sit. Furthermore, all the poles of the approximant are simple and located on the negative real axis, {\it{i.e.}}, on the physical axis, and residues are positive.

Therefore, from a physical point of view, the Pad\'e approximant to (\ref{green}) can be built, at least formally, out of the infinite number of terms of the chiral expansion. Pad\'e poles can then be interpreted as meson mass-poles whereas Pad\'e residues play the r\^ole of the meson decay constants. Convergence then means that our Pad\'e approximant has to match asymptotically the real QCD spectrum.

The strategy followed in \cite{Migdal} was to construct the symmetric $[N,N]$ approximant around an expansion point $q^2=-\mu^2$ on the far Euclidean axis. The Pad\'e equations then read \cite{Migdal,Erlich}
\begin{equation}\label{padeeq.} 
\fr{d^n}{d(q^2)^n}\bigg[\Pi_V(q^2)\Qcal_N(q^2)-\Pcal_N(q^2)\bigg]\bigg|_{q^2=-\mu^2}=0,\qquad n=0,...,2N \, .
\end{equation}
It is worth noting at this point that the convergence theorems above no longer apply if one builds the Pad\'e away from the origin. Interesting exceptions are meromorphic functions. There it can be proved that convergence is guaranteed as long as the Pad\'e is constructed on any non-singular point of the real $q^2$ axis.\footnote{I thank G.~L\'opez-Lagomasino for helping me clarify this issue.} In the strict large-$N_c$ limit Green's functions are known to be meromorphic, which means, in particular, that with the model of section 2 the Pad\'e constructed around an arbitrary $q^2=-\mu^2$ converges. However, for finite $N_c$ nothing compels our Pad\'e to approach the original function. It is true that often Pad\'e approximants converge even when there is no theorem to guarantee their convergence, but this pragmatic approach is far from being rigorous.

Here we are just interested in the mass spectrum, which is encoded in $\Qcal_N(q^2)$. We will therefore keep only the upper half of the previous set of equations, where the contribution of $\Pcal_N(q^2)$ vanishes, namely
\begin{equation}
\fr{d^n}{d(q^2)^n}\bigg[\Pi_V(q^2)\Qcal_N(q^2)\bigg]\bigg|_{q^2=-\mu^2}=0,\qquad n=N+1,...,2N\, . 
\end{equation}
Use of Cauchy's integral formula leads to
\begin{equation}\label{Cauchy}
\fr{d^n}{d(q^2)^n}\bigg[\Pi_V(q^2)\Qcal_N(q^2)\bigg]\bigg|_{q^2=-\mu^2}=\fr{n!}{2\pi i}\oint_{\gamma}\fr{dq^2}{(q^2+\mu^2)^{n+1}}\bigg[\Pi_V(q^2)\Qcal_N(q^2)\bigg]\, ,
\end{equation}
where $\gamma$ is the contour depicted in figure 1. We can relate the previous expression to the spectral function by using the conventional strategy in sum rules of splitting the contour in two paths (see, for instance, \cite{Rafael}). In $N_c=3$ QCD one can readily show that the integral over the circle $|q^2|=s_0$ vanishes when the contour is sent to infinity, and only the integral over the physical axis contributes. This is related to the fact that the asymptotic behavior of the spectral function goes to a constant and so\footnote{In the strict large-$N_c$ limit, equation (\ref{asym}) is ill-defined because of the discrete nature of the spectral function. However, one can show that (\ref{Kallen}) still holds. See the Appendix for a detailed proof.}
\begin{equation}\label{asym}
\lim_{t\rightarrow \infty}\fr{1}{\pi}{\mathrm{Im}}\,\Pi_V(t)\,\Qcal_N(t)\sim t^N\, .
\end{equation}
Therefore, $N+1$ derivatives are sufficient to get a K\"all\'en-Lehmann representation for the Pad\'e equations, namely
\begin{eqnarray}\label{Kallen}
\fr{d^n}{d(q^2)^n}\bigg[\Pi_V(q^2)\Qcal_N(q^2)\bigg]\bigg|_{q^2=-\mu^2}&=&n!\int_0^{\infty}\fr{dt}{(t+\mu^2)^{n+1}}\bigg[\fr{1}{\pi}{\mathrm{Im}}\,\Pi_V(t)\Qcal_N(t)\bigg]\, ,\\
&&\qquad \qquad \qquad \qquad \qquad n=N+1, ... ,2N\, ,\nonumber
\end{eqnarray} 
where we made use of Schwarz reflection formula and the reality of $\Qcal_N(q^2)$.
\begin{figure}
\renewcommand{\captionfont}{\small \it}
\renewcommand{\captionlabelfont}{\small \it}
\centering 
\psfrag{A}{${\mathrm{Re}}\, q^2$}
\psfrag{B}{$\gamma$}
\psfrag{C}{\footnotesize{$(q^2=-\mu^2)$}}
\includegraphics[width=2.5in]{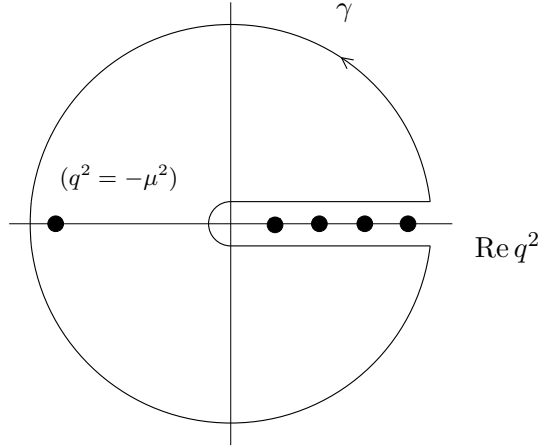}
\caption{Contour integration chosen in (\ref{Cauchy}). The singularities on the right half-plane are the poles of $\Pi_V(q^2)$, sitting at $q^2=m_V^2+ak$ for natural $k$. The pole on the left half-plane is the pole with multiplicity $(n+1)$ explicitly shown in the denominator of (\ref{Cauchy}). The circle is defined by $|q^2|=s_0$, which will be eventually sent to infinity.}\label{fig:plot}
\end{figure}
Following Migdal's strategy, at that point we will rely on (local) quark-hadron duality and assume that the spectral function can be inferred from the OPE of (\ref{ope}). Then,
\begin{equation}\label{naive}
\fr{1}{\pi}{\mathrm{Im}}\,\Pi_V(t)\simeq \fr{4}{3}\fr{N_c}{(4\pi)^2}\theta(t)+\sum_{k=0}^{\infty}c_{(2k+2)}\,\delta^{(k)}(t)\, ,
\end{equation}
where $\delta^{(k)}(t)$ stands for the $k$-th derivative of the Dirac delta and $\delta^{(0)}(t)\equiv \delta (t)$. If we just keep the constant term in (\ref{naive}) --{\it{i.e.}}, the logarithmic piece in (\ref{ope})--, then the Pad\'e equations (\ref{Kallen}) reduce to
\begin{equation}\label{pad}
\int_0^{\infty}\fr{dt}{(t+\mu^2)^{n+1}}\Qcal_N(t)=0,\qquad n=N+1, ... , 2N\, .
\end{equation} 
By reorganizing the powers of $(t+\mu^2)$ in (\ref{pad}), Migdal realized that $\Qcal_N(q^2)$ are proportional to Jacobi polynomials. Actually, it seems that the first derivation of the Pad\'e approximant to the logarithm was performed by Jacobi, Rouch\'e and Gauss.\footnote{See \cite{Weideman} for details.} Their result for the poles was 
\begin{equation}\label{result}
\Qcal_N(q^2)=_{\,\,2}\!\!F_1\left(-N,-N;1;-\fr{q^2}{\mu^2}\right)=(q^2+\mu^2)^N\,P_N^{(0,0)}\left(\fr{\mu^2-q^2}{\mu^2+q^2}\right)\, ,
\end{equation}
where ${}_{\,2}F_1(a,b;c;d)$ is Gauss' hypergeometric function and $P_N^{(0,0)}$ stands for the $(0,0)$-Jacobi polynomial, {\it{i.e.}}, the Legendre polynomial. Up to an overall constant out front, this last expression is the one found in the recent analysis of \cite{Erlich}. Plugging (\ref{result}) in (\ref{padeeq.}) we can solve for ${\cal{P}}_N(q^2)$. Assembling all the pieces, the final result for the correlator reads
\begin{equation}\label{padelog}
\Pi_V(q^2)\simeq \frac{2}{(q^2+\mu^2)^N}\,\sum_{k=0}^{N}\left (\begin{array}{c}
k \\
j
\end{array}\right)^2 \left[\frac{H_{N-k}-H_k}{P_N^{(0,0)}(\chi)}\right]\left(-\fr{q^2}{\mu^2}\right)^k\, , \qquad \chi=\frac{\mu^2-q^2}{\mu^2+q^2}\, ,
\end{equation}
where $H_k$ are the harmonic numbers. The expression above is the Pad\'e approximant to the logarithm, {\it{i.e.}}, when $N \rightarrow \infty$ the logarithm we started from is recovered. 

\section{Migdal's limit and quark-hadron duality}

However, Migdal went one step beyond the strict Pad\'e approximation. At low energies ($q^2\ll\mu^2$) the argument of the Jacobi polynomial above can be Taylor-expanded to yield
\begin{equation}
\lim_{q^2\rightarrow 0}\fr{\mu^2-q^2}{\mu^2+q^2}=\left(1-2\fr{q^2}{\mu^2}\right)\, .
\end{equation}
Then, using the asymptotic formula \cite{Abramowitz}
\begin{equation}
\lim_{N\rightarrow \infty} P_N^{(0,b)}\left(1-\fr{\xi^2}{2N^2}\right)=J_0(\xi)\, ,
\end{equation}
one ends up with
\begin{equation}\label{solution1}
\Qcal_N(q^2)=J_0\left(2N\sqrt{\fr{q^2}{\mu^2}}\,\,\right)\, ,
\end{equation}
and the full correlator takes the form
\begin{equation}\label{migpade}
\Pi_V(q^2)=-\frac{4}{3}\frac{N_c}{(4\pi)^2}\left[\log\frac{q^2}{\mu^2} -\pi\frac{Y_0\left(q\Lambda\,\,\right)}{J_0\left(q\Lambda\,\,\right)}\right]\, , \qquad \Lambda=\frac{2N}{\mu}\, .
\end{equation}
It is interesting to examine the last expression both in the Euclidean and Minkowki regions. In the far Euclidean it can be shown that the correlator reduces to ($Q^2=-q^2$)
\begin{equation}\label{inst}
\Pi_V(Q^2)=-\frac{4}{3}\frac{N_c}{(4\pi)^2}\log\frac{Q^2}{\mu^2}+{\cal{O}}(e^{-2Q\Lambda})\, ,\qquad (Q^2\gg 0).
\end{equation}
The last term does not stem from the Pad\'e approximation. Rather, it is a consequence of the combined $q^2\ll\mu^2$ and $N\rightarrow \infty$ limits taken above. Despite its exponential suppression, the new term is of the outmost importance. First and foremost, it does not belong to the OPE, which is a series expansion in inverse powers of $Q^2$. The presence of exponentially suppressed terms in the Euclidean half-plane is a reminder that the OPE is not an allowed expansion in the whole $q^2$ complex plane. At least on the physical axis the expansion breaks down. Generically one writes
\begin{equation}\label{analcont}
\Pi_V(q^2)\simeq \Pi_V^{OPE}(q^2)+\Delta(q^2)\, ,\qquad (q^2\gg 0),
\end{equation}
where $\Delta(q^2)$ is referred to as {\it{quark-hadron duality violation}}. This extra term can be understood on the basis that the OPE, being a regular expansion, loses track of the precise details of the spectrum, which are then encoded in the $\Delta(q^2)$ function. $\Delta(q^2)$ thus contains the singularities (cuts and poles) of $\Pi_V(q^2)$. Even though the origin of such extra term is well understood, its characterization is a most challenging task. 

Two basic patterns have been studied so far in the literature, associated with infinite-size and finite-size singularities \cite{Shifman}. In the former case, based on a model of resonances, the authors of \cite{Cata05paper,Cata05} concluded that
\begin{equation}\label{dual}
\Delta(q^2)\simeq {\cal{H}}(q^2)\,e^{-|q^2|{\cal{F}}(\phi,N_c)}\, .
\end{equation}
where ${\cal{H}}(q^2)$ and ${\cal{F}}(\phi,N_c)$ are in general unknown. The only generic constraint comes from the fact that in the strict large-$N_c$ limit the spectrum is an array of Dirac deltas. This requires that ${\cal{F}}(\phi,N_c\rightarrow \infty)$ has to cancel on the Minkowki axis ($\phi=0,2\pi$), therefore killing the exponential damping of Eq.(\ref{dual}).

Finite-size singularities were studied in an instanton model \cite{Shifman}. Quite unexpectedly, the construction of Migdal, even though it is claimed to be a model of resonances in the large-$N_c$ limit, leads to an instanton-like pattern for duality violations, as first noted in \cite{Hirn}.\footnote{It is worth noting that instanton-like duality violations are expected to be suppressed at large-$N_c$, as Migdal himself noted. Duality breakdown due to resonances, however, is a leading order effect in the $1/N_c$ expansion.}

 Obviously, the pattern is induced by the appearance of the infrared regulator $\Lambda$, which arises after the limits $q^2\ll\mu^2$, $N\rightarrow \infty$ have been taken. It is important to notice that both limits are correlated, {\it{i.e.}}, $\mu/N$ is to stay constant. This peculiar continuum limit spoils the convergence of the Pad\'e approximant. The best way to visualize this is to consider the expression for the Pad\'e residue
\begin{equation}
\Pi_V(q^2)\Qcal_N(q^2)-\Pcal_N(q^2)={\cal{R}}_N(q^2)\, ,
\end{equation}
in terms of which our correlator can be expressed as
\begin{equation}\label{resi}
\Pi_V(q^2)=\frac{\Pcal_N(q^2)}{\Qcal_N(q^2)}=-\frac{4}{3}\frac{N_c}{(4\pi)^2}\log\frac{-q^2}{\mu^2}-\frac{{\cal{R}}_N(q^2)}{\Qcal_N(q^2)}\, .
\end{equation}
For the logarithm it can be shown that the last term above is a Meijer G function which vanishes in the continuum limit ($N\rightarrow \infty$), thus ensuring that our Pad\'e approximant is indeed convergent. Notice, however, that fixing the ratio $\mu/N$ gives a non-vanishing residue, {\it{cf.}} Eq.(\ref{migpade}). Convergence of the Pad\'e approximant is lost due to a source of duality violation put by hand and parametrized in terms of an infrared scale $\Lambda$.

Let us now turn our attention to the Minkowski axis to analize the impact of the new scale $\Lambda$. According to our previous discussion about the duality violating pieces, they have to account for the singularities of the correlator. This is indeed what we observe for the toy model of section 2. The asymptotic expansion of Eq.(\ref{dig}) leading to the OPE of Eq.(\ref{ope}) is no longer valid in the physical axis, where the digamma function has its poles. This can be circumvented by recalling the reflection property of the digamma function,
\begin{equation}\label{reflection}
\psi(\xi)=\psi(-\xi)-\pi\cot{\pi\xi}-\fr{1}{\xi}\, ,
\end{equation}
in order to be able to analytically-continue the OPE to the Minkowski region. Therefore, in our specific toy model, duality violations can be computed analytically. Matching onto (\ref{analcont}), one gets
\begin{equation}\label{duali}
\Delta(q^2)=\fr{\pi F^2}{a} \cot{\bigg[\pi\fr{-q^2+m_V^2}{a}\bigg]}\, .
\end{equation}
Recall that indeed the cotangent encodes information about the poles of the digamma function, something that the OPE fails to reproduce. Poles are equidistantly located on the real axis and are given by the expression
\begin{equation}\label{cotpoles}
\pi\cot{\pi\xi}=\fr{1}{\xi}+2\xi\sum_{n=1}^{\infty}\fr{1}{\xi^2-n^2}\, .
\end{equation}
A similar thing happens with Eq.(\ref{migpade}). The analytic continuation can be done straightforwardly and
\begin{equation}
\Delta(q^2)=\frac{4}{3}\frac{N_c}{(4\pi)^2}\left[\pi\frac{Y_0\left(q\Lambda\,\,\right)}{J_0\left(q\Lambda\,\,\right)}\right]\, .
\end{equation}
The seemingly negligible exponential terms in the far Euclidean are now poles sitting at zeros of the Bessel $J_0$ function. In the Pad\'e approximant, when $N\rightarrow \infty$ the
residue ${\cal{R}}_N(q^2)$ vanishes and we recover the logarithm. In Migdal's
approach, in contrast, the poles do not merge into the logarithmic cut.  

Interestingly, both the toy model of section 2 and Migdal's model have 5-dimensional duals in the context of the gauge/gravity duality, the so-called soft-wall and hard-wall models, respectively. It seems interesting to briefly comment on how what we have discussed carries over to the fifth dimension.    

\section{Connection with holographic QCD models}
In this section we will pursue the relationship between QCD and holographic QCD duals suggested in \cite{Erlich}. If Migdal's original work is related to hard-wall holographic models, there should be a similar relationship between our toy model and holographic duals with built-in Regge behavior. This relationship was recently found in \cite{Karch}.     

The 5-dimensional model they started from can be written as
\begin{equation}\label{action}
S=-\int\,d^4x\,dz\,e^{-\Phi(z)}\sqrt{g}\fr{1}{4g_5^2}{\mathrm{Tr}}\,\bigg[(F_{{\hat{\mu}}{\hat{\nu}}}F^{{\hat{\mu}}{\hat{\nu}}})_L+(F_{{\hat{\mu}}{\hat{\nu}}}F^{{\hat{\mu}}{\hat{\nu}}})_R\bigg]\, ,
\end{equation}
where the hatted indices ${\hat{\mu}},{\hat{\nu}}=0,1,2,3,4$; the metric is parametrised as
\begin{equation}
g_{\hat{\mu}\hat{\nu}}\,dx^{\hat{\mu}}dx^{\hat{\nu}}=e^{2A(z)}(\eta_{\mu\nu}dx^{\mu}dx^{\nu}+dz^2)\, ,\qquad \eta_{\mu\nu}={\mathrm{diag}}(-1,1,1,1)\, ,
\end{equation}
and $\Phi(z)$ stands for the dilaton field. For our purposes, namely the spectrum of vector mesons, additional bulk fields in the action (\ref{action}) are irrelevant. 

The eigenvalue equation for the vectorial modes can be expressed as a Sturm-Liouville differential equation
\begin{equation}
\fr{d}{dz}\left(e^{-B(z)}\fr{d}{dz}V_n\right)+m_n^2\,e^{-B(z)}V_n=0\, ,
\end{equation}
where $B(z)=\Phi(z)-A(z)$. After a change of variable, the previous equation can be recast in the Schr\"odinger-like form
\begin{equation}\label{schro}
\left\{-\fr{d^2}{dz^2}+U(z)\right\}\psi_n(z)=m_n^2\,\psi_n(z),\qquad U(z)=\left(\fr{B^{\prime}}{2}\right)^2-\fr{B^{\prime\prime}}{2}\, .
\end{equation}
In the so-called hard-wall models, $A=-\log{z}$, such that the metric is AdS, and the dilaton is taken to be constant, $\Phi(z)=\phi,\,\epsilon\leq z\leq \Lambda$, where $z=\epsilon$ and $z=\Lambda$ are the four-dimensional boundaries. Then the potential is given by
\begin{equation}
U(z)= \fr{3}{4z^2}\, .
\end{equation}
The Schr\"odinger equation has Bessel functions as solutions, and the correlator takes the form of Eq.(\ref{migpade}), yielding $m_n^2\sim n^2$. The emergence of the scale $\Lambda$ is, in the context of holographic duals, due to the presence of the infrared boundary brane, which breaks conformal invariance and leads to the terms at the outmost right of Eq. (\ref{migpade}). Duality violations are therefore provided by the infrared boundary conditions, and its pattern as a finite-size singularity has now a transparent geometrical interpretation: it has to do with the finite size of the fifth dimension.

However, with the geometrical insight of the holographic models we can go further. The fact that the eigenfunctions are the Bessel $J_0$ function means that their overlap is non-zero. In the language of holography, this leads to non-vanishing decay widths for the resonances. Therefore, back to 4 dimensions, the correlated limit of Migdal, which swallows the degree $N$ of the Legendre polynomials into the argument of the Bessel $J_0$ function, breaks down orthogonality and gives an instanton-like size to the resonances. 

All these shortcomings disappear if there is no infrared boundary brane. Then there is no instanton-like duality breaking. Moreover, orthogonality of the wave-functions is left unbroken and resonances are infinitely narrow, as one expects from a large-$N_c$ model of resonances. Furthermore, if one modifies the infrared behavior of the model without an infrared brane, one can reproduce Regge scaling. This last point was the original motivation of the authors of \cite{Karch}. According to them, the requirement of Regge behavior leads to a (asymptotically) unique choice of the dilaton background, namely $\Phi(z)\sim cz^2$. In particular, the simplest potential incorporating the right leading order QCD short distances and Regge behavior takes the form
\begin{equation}\label{laguerre}
U(z)= c^2z^2+\fr{3}{4z^2}\, ,
\end{equation}    
which corresponds to the three-dimensional harmonic oscillator in quantum mechanics. The harmonic term is the dilaton contribution, whereas the centrifugal barrier term is due to the AdS metric. The Schr\"odinger equation is solved through Laguerre polynomials \cite{Karch}
\begin{equation}
\psi_n(z)\sim e^{-\frac{1}{2}cz^2}z^{3/2}L_n^1\,(cz^2)\, ,
\end{equation}
with the spectrum scaling as
\begin{equation}\label{mass}
m_n^2=4cn+4c\, ,  \qquad n=0,1,...
\end{equation} 
Furthermore, the decay constants can be computed to give a constant,
\begin{equation}\label{decay}
F_n^2=\fr{2}{g_5^2}c\equiv F^2\, .
\end{equation}
As noted by \cite{Karch}, comparison of Eqs. (\ref{mass}) and (\ref{decay}) with Eq. (\ref{regge}) shows that the 5-dimensional model with linear confinement is dual to the toy model of section 2 with $a=4c$ and $m_V^2=4c$. This immediately means that the dilaton quadratic profile is the holographic dual to the duality violating piece
\begin{equation}
\Delta(q^2)=\frac{\pi}{2g_5^2}\cot{\left[\pi\frac{-q^2+4c}{4c}\right]}\, .
\end{equation}

From a phenomenological point of view, this soft-wall model has two free parameters, $g_5$ and $c$. $g_5$ can be determined by matching the model to the parton model logarithm, while $c$ can be extracted by imposing that $m_V^2=m_{\rho}^2$.  Taking $m_{\rho}=776$ MeV, we find
\begin{equation}
g_5^2=\frac{12\pi^2}{N_c}\, , \qquad c=\frac{m_{\rho}^2}{4}\simeq 150\,{\mathrm{MeV}}^2\, .
\end{equation}
It is interesting to notice that the ratio $m_V^2/a=1$ is fixed in the 5-dimensional soft-wall model, as opposed to our toy model, where $a$ and $m_V^2$ are free parameters. In particular this means that the soft-wall model gives a non-zero negative dimension-two condensate. Using the OPE expansion of Eq.(\ref{cond}) one finds
\begin{equation}
c_2=-F^2\left(\frac{m_V^2}{a}-\frac{1}{2}\right)=-\frac{1}{2\pi^2}\, .
\end{equation}
The model also underestimates the Regge slope, giving $a=0.6$ GeV$^2$ while the phenomenological one hovers around $a\simeq 1.2$ GeV$^2$. Finally, the decay constant is predicted to be $F=123$ MeV, while its experimental value lies at $F=210$ MeV. Interestingly, both the Regge slope and the decay constant adjust to their experimental values with the choice $m_V^2/a=2$. Notice that this ratio also cancels the dimension-two condensate.

It would therefore be really interesting to know if it is possible to modify the soft-wall model so that $m_V^2/a=2$.

\section{Conclusions}

Upon reassessing Migdal's work we have shown that, even though the starting point is a well-defined Pad\'e approximant to the parton model logarithm, the correlated limits taken thereafter not only break the convergence of the approximant but also modify the analytic structure of the correlator. We have seen that the correlated limits, which lead to the eventual emergence of an infrared scale $\Lambda$, are nothing but a modeling of quark-hadron duality violations. However, this modeling corresponds not to what one expects from a model of resonances in large-$N_c$ but to finite-size effects, most commonly associated to instantons. Actually, the nature of the resonances resembles more that of instanton-like excitations rather than quark-antiquark bound states.

It is worth stressing that in Migdal's approach it is the duality violating piece what brings information on the spectrum. This is a general feature: the OPE, being a regular expansion, is completely blind to the singularities in the $q^2$ plane, and therefore unable to reproduce the details of the spectrum. Those singularities are encoded in the form of duality breakdown. Therefore, contrary to what was conveyed in Migdal's work, the addition of ultraviolet corrections to the correlator does not help in determining the spectrum. This is related to the well-known fact that different functions can lead to the same asymptotic behavior. As an academic exercise, consider our toy model of large-$N_c$ QCD with built-in Regge behavior. We can fine-tune it such that it matches the first $k$ OPE vacuum condensates.\footnote{This can be easily achieved, {\it{e.g.}}, by relaxing our Eq.(\ref{regge}) so that the first $k$ decay constants $F_n$ are left as free parameters.} Let us imagine that all OPE condensates can be accounted for this way. Therefore, the Pad\'e approximant (and Migdal's construction upon it) is now unable to distinguish our toy model from real QCD. From perturbative QCD and, most generally, the OPE, the hadronic spectrum cannot be reconstructed. This same statement was already made in a slightly different context in \cite{Cata06}.  

Migdal's original motivation for taking the correlated continuum limit was to make quark-hadron duality more accurate. We have shown that it is precisely the other way round: quark-hadron duality was enforced from the beginning and preserved by the Pad\'e approximant. Duality breakdown only occurred after the correlated limits were taken, generating the exponentially suppressed terms of Eq.(\ref{inst}). 

Even from a conceptual point of view, in real QCD duality is expected to be restored at high energies rather than at low energies. In $N_c=3$ QCD, and even at large but finite $N_c$, resonances in the spectrum develop a width, expected to be proportional to 
\begin{equation}\label{width}
\Gamma_V(n)\sim \fr{\sqrt{n}}{N_c}\, ,
\end{equation}
$n$ being the excitation number. Therefore, increasing widths of resonances makes them overlap and eventually the spectrum is smeared out into a continuum: local duality sets in. This means that at sufficiently large values of momentum $t$ on Minkowski axis the duality picture has to be very accurate.\footnote{This is the rationale behind the sum rule program.} Actually, based on a model, the authors of \cite{Blok} found
\begin{equation}
{\mathrm{Im}}\,\Delta(t)\sim {\mathrm{exp}}\bigg(-\fr{\alpha t}{N_c}\bigg)\cos{\beta t}
\end{equation}  
as a particular realization of Eq.(\ref{dual}), $\alpha$ and $\beta$ being constants. However, in the strict large-$N_c$ limit $N_c$ is sent to infinity from the very beginning. Resonances therefore stay infinitely narrow no matter the excitation number and local duality never sets in.  

The large-$N_c$ and low-$q^2$ limits taken by Migdal avoid the continuum region of QCD, precisely where the assumption of duality is allowed. At low energies resonance poles are well-defined only because duality violations are sizeable.  In large-$N_c$ duality breaks down by construction. If we attempt to reconstruct the spectrum of QCD, the assumption of duality in Migdal's work is inherently inconsistent. 

The only consistent way to make contact with hadronic resonances with a Pad\'e approximant would be to incorporate duality-violating terms in the spectral function before building the Pad\'e approximant. Notice, however, that this does not yield any predictions: knowledge of the spectrum is already required prior to the Pad\'e construction. Since duality breakdown has to be inserted by hand, in model building it is much better to start from an ansatz for the spectral function and build the correlator using dispersion relations \cite{Cata05paper}, if we are working in 4 dimensions from the start, or playing with the infrared properties of an holographic model, if we use 5-dimensional models in AdS backgrounds.

Gauge/gravity duals provide a more intuitive physical picture of quark-hadron duality breakdown, based on geometry. The somewhat misterious infrared scale in Migdal's model can now be interpreted as the infrared boundary brane. The presence of this brane breaks conformal invariance and generates the terms that we identified as quark-hadron duality violations. Both Migdal's model and the hard-wall model start from the parton model logarithm and model duality breakdown in exactly the same way. Therefore, the shortcomings mentioned for Migdal's model carry over to the hard-wall model in 5 dimensions. However, in gauge/gravity duals there is a way to get rid of those shortcomings by removing the infrared brane and changing the infrared properties of the metric and/or the dilaton field at the same time. This corresponds to a change in the modeling of the duality breakdown terms. If this is done smartly, one can end up with a model with built-in Regge trajectories \cite{Karch}. A quadratic dilaton background then leads to a dual version of our toy model of section 2. Unfortunately, the fact that the ratio between $m_{\rho}$ and the Regge slope in the soft-wall model are fixed leads to poor phenomenological predictions. 

The statement that holographic models were leading in an essential way to non-Regge behavior \cite{Shifman1} was based on the belief that the metric in the ultraviolet was determining the spectrum. This is essentially the same philosophy lying behind the approach developed by Migdal. In holographic models, as in 4-dimensions, this is not true. Notice that one could get rid of the centrifugal barrier in Eq.(\ref{laguerre}) and work in flat space-time with the same dilaton background. By solving the corresponding Schr\"odinger equation the eigenfunctions would be proportional to Hermite polynomials, and its scaling would still be Regge-like, $m_n^2\sim n$. Obviously, the interplay of the metric and the dilaton would give rise to a different background and this will affect, {\it{e.g.}}, the determination of decay constants and the slope of the spectrum (the $a$ parameter in our toy model) but, importantly, will not change its scaling properties. Therefore, it is not the ultraviolet of the theory but some modeling of quark-hadron duality breakdown what is necessary to model the spectrum. 

This brings up the interesting question of whether there is hope in the AdS-QCD models to build the holographic dual of QCD. Unfortunately, very little is known about the form of duality violations in real QCD.\footnote{See \cite{Shifman,Cata05paper,Blok} for some recent studies in the large-$N_c$ limit.} In order to achieve the goal successfully it seems that a better knowledge of non-perturbative QCD is mandatory.



\section*{Acknowledgements}

I am grateful to M.~Piai, D.~T.~Son, M.~Unsal, A.~Karch, V.~Sanz and M.~P\'{e}rez-Victoria for useful conversations and to C.~P.~Herzog, S.~Peris and M.~J.~Strassler for their critical reading of the manuscript and the suggestions that followed. I also would like to acknowledge very fruitful correspondence with S.~J.~Brodsky, G.~F.~de T\'eramond, J.~Hirn, G.~L\'opez-Lagomasino, I.~Low and E.~J.~Weniger. My thanks to the Department of Physics of the University of Washington at Seattle, where this work has been carried out, for their hospitality. The author is supported by the Fulbright Program and the Spanish Ministry of Education and Science under grant no. FU2005-0791.

\section*{Appendix}
As stated in the main text, the vanishing of contour integrals in two-point functions can be inferred from the asymptotic behavior of the spectral function, leading in our case to 
\begin{equation}
\lim_{t\rightarrow \infty}\fr{1}{\pi}{\mathrm{Im}}\,\Pi_V(t)\Qcal_N(t)\sim t^N\, .
\end{equation}
However, in the large-$N_c$ limit the above equation is strictly speaking 
ill-defined. Therefore, one has to adopt a slightly different strategy. 
We start from
\begin{eqnarray}\label{Cauchy1}
\fr{d^n}{d(q^2)^n}\bigg[\Pi_V(q^2)\Qcal_N(q^2)\bigg]\bigg|_{q^2=-\mu^2}&=&\fr{n!}{2\pi i}\oint_{\gamma}\fr{dq^2}{(q^2+\mu^2)^{n+1}}\bigg[\Pi_V(q^2)\Qcal_N(q^2)\bigg]=\nonumber\\
&=&\fr{n!}{2\pi i}\oint_{|q^2|=s_0}\,\,\,\,\fr{dq^2}{(q^2+\mu^2)^{n+1}}\bigg[\Pi_V(q^2)\Qcal_N(q^2)\bigg]+\nonumber\\
&+&n!\int_0^{s_0}\fr{dt}{(t+\mu^2)^{n+1}}\bigg[\fr{1}{\pi}{\mathrm{Im}}\,\Pi_V(t)\Qcal_N(t)\bigg]\, .
\end{eqnarray}
where the limit $s_0\rightarrow \infty$ is to be understood henceforth. To get to (\ref{Kallen}) in the main text we have to show that the integral in the second line above identically vanishes. Even though we will carry all explicit calculations with the model of section 1, we will find that this statement is a general result of large-$N_c$ QCD itself.

Our model reads
\begin{eqnarray}
\Pi_V(q^2)&=&\fr{F^2}{a}\left[\psi\left(\fr{m_V^2}{a}\right)-\psi\left(\fr{-q^2+m_V^2}{a}\right)\right]\, ,\nonumber\\
\fr{1}{\pi}{\mathrm{Im}}\,\,\Pi_V(t)&=&F^2\sum_{n=0}^{\infty}\delta(t-m_V^2-an)\, .
\end{eqnarray}
Recall that the digamma function $\psi(\xi)$ is analytic everywhere in the complex $\xi$ plane except for simple poles at negative integer values of the argument, {\it{i.e.}}, $\xi=0,-1,-2,...$. Moreover, it satisfies 
\begin{equation}
{\mathrm{Res}}\,[\psi(\xi);\xi]=-1,\qquad \xi=0,-1,-2,...
\end{equation}
The second line in (\ref{Cauchy1}) can then be straightforwardly evaluated by reiterative application of Cauchy's residue theorem. The integrand has a pole of order $n$+1 in the Euclidean axis sitting at $q^2=-\mu^2$ along with an infinite number of equidistant (simple) poles in the Minkowski axis, at
\begin{equation}
q^2=m_V^2+ak,\qquad k=0,1,...
\end{equation} 
Application of the residue theorem then yields
\begin{eqnarray}\label{first}
\fr{n!}{2\pi i}\oint_{|q^2|=s_0}\,\,\fr{dq^2}{(q^2+\mu^2)^{n+1}}\bigg[\Pi_V(q^2)\Qcal_N(q^2)\bigg]
&=&\lim_{q^2\rightarrow -\mu^2}\fr{d^n}{d(q^2)^n}\bigg[\Pi_V(q^2)\Qcal_N(q^2)\bigg]+\nonumber\\
&\!\!\!\!\!\!\!\!\!\!\!\!\!\!\!\!\!\!\!\!\!\!\!\!\!\!\!\!\!\!\!\!\!\!\!\!\!\!\!\!\!\!\!\!\!\!\!\!\!\!\!\!\!\!\!\!\!\!\!\!\!\!\!\!\!\!\!\!\!\!\!\!+&\!\!\!\!\!\!\!\!\!\!\!\!\!\!\!\!\!\!\!\!\!\!\!\!\!\!\!\!\!\!\!\!\!\!\!\!\!\!\!\!\!\!\!\!\lim_{q^2\rightarrow m_V^2+ak}n!\sum_{k=0}^{\infty}\bigg[\fr{q^2-(m_V^2+ak)}{(q^2+\mu^2)^{n+1}} \Pi_V(q^2)\Qcal_N(q^2) \bigg]\, .
\end{eqnarray}
Obviously, the first term is the left-hand side we started from in (\ref{Cauchy1}). Therefore, the second line above has to cancel the integral over the spectral function in (\ref{Cauchy1}). Indeed,
\begin{equation}\label{disp}
\int_0^{s_0}\fr{dt}{(t+\mu^2)^{n+1}}\bigg[\fr{1}{\pi}{\mathrm{Im}}\,\Pi_V(t)\Qcal_N(t)\bigg]=F^2\sum_{k=0}^{\infty}\fr{\Qcal_N(m_V^2+ak)}{(m_V^2+ak+\mu^2)^{n+1}}\, ,
\end{equation}  
whereas
\begin{eqnarray}\label{firstsol}
\lim_{q^2\rightarrow m_V^2+ak}&&\!\!\!\!\!\!\!\!\!\!\!\!\sum_{k=0}^{\infty}\bigg[\fr{q^2-(m_V^2+ak)}{(q^2+\mu^2)^{n+1}} \Pi_V(q^2)\Qcal_N(q^2) \bigg]=\nonumber\\
&=&\!\!\!\!\!\lim_{q^2\rightarrow m_V^2+ak}F^2\sum_{k=0}^{\infty}\bigg[\left(\fr{-q^2+m_V^2}{a}+k\right) \psi\left(\fr{-q^2+m_V^2}{a}\right)\fr{\Qcal_N(q^2)}{(q^2+\mu^2)^{n+1}} \bigg]\, ,
\end{eqnarray}
which equals (\ref{disp}) but for a change in sign.\footnote{This becomes obvious if one deforms the path $\gamma$ such it does not include the pole in the Euclidean. Then Cauchy's integral theorem asserts that the integral over the circle equals that over the physical axis.}

Now we only need to show that the first line in (\ref{first}) cancels (\ref{firstsol}), or equivalently, as we have just checked, equals (\ref{disp}). In other words, the contribution from the infinite poles on the Minkowski axis counterplays that of the pole at the Euclidean axis. Applying the binomial formula to the first line of (\ref{first}), one finds
\begin{eqnarray}\label{last}
\fr{d^n}{d(q^2)^n}&&\!\!\!\!\!\!\!\!\!\!\!\!\!\!\!\!\!\bigg[-\fr{F^2}{a}\psi\left(\fr{-q^2+m_V^2}{a}\right)\Qcal_N(q^2)\bigg]\bigg|_{q^2=-\mu^2}=\nonumber\\
&=&-\fr{F^2}{a}\sum_{j=1}^n\left (\begin{array}{c}
n \nonumber\\
j
\end{array}\right)
\left(\fr{-1}{a}\right)^j\bigg[\psi_j\left(\fr{-q^2+m_V^2}{a}\right)\Qcal_N^{(n-j)}(q^2)\bigg]\bigg|_{q^2=-\mu^2}=\nonumber\\
&=&F^2\sum_{k=0}^{\infty}\sum_{j=1}^{n}\fr{n!}{(n-j)!}\fr{\Qcal_N^{(n-j)}(-\mu^2)}{(m_V^2+ak+\mu^2)^{j+1}}\, ,
\end{eqnarray}
where in the last line we have used the definition of the polygamma function, namely
\begin{equation}
\psi_j(\xi)=\fr{d^j}{d\xi^j} \psi(\xi)=(-1)^{j+1}j!\sum_{k=0}^{\infty}\fr{1}{(\xi+k)^{j+1}}\, .
\end{equation}
The last line in (\ref{last}) can be recast in the more useful way 
\begin{equation}
\sum_{k=0}^{\infty}\fr{1}{(m_V^2+ak+\mu^2)^{n+1}}\sum_{j=1}^n\fr{n!}{(n-j)!}\Qcal_N^{(n-j)}(-\mu^2)(m_V^2+ak+\mu^2)^{n-j}\, .
\end{equation}
Therefore, in order to match (\ref{disp}), it follows that
\begin{equation}
\Qcal_N(m_V^2+ak)=\sum_{j=1}^n\fr{1}{(n-j)!}\Qcal_N^{(n-j)}(-\mu^2)(m_V^2+ak+\mu^2)^{n-j}\, ,
\end{equation} 
but the sum over $j$ is just the Taylor expansion of $\Qcal_N(q^2)$ around $q^2=-\mu^2$, evaluated at $q^2=m_V^2+an$, and truncated after $n+1$ terms. However, being $\Qcal_N(q^2)$ a polynomial of degree $N$, the previous relation is exact if $n>N$, which is one of our starting points. This completes the proof.

Recall that the proof, as anticipated, relies only upon the analytic properties of two-point Green's functions in large-$N_c$, namely the existence of no cuts and just simple poles. Therefore, the result is model-independent and valid for large-$N_c$ QCD.

\end{document}